\documentclass{aa}
\usepackage{graphicx,natbib}
\def\bp{$\beta$\,Picto\-ris}

\def\cs{cir\-cum\-stel\-lar}

\def\h{\hfill\break}

\newcommand{\dma}[1]{_{\mathrm{#1}}}
\begin{document}
\title{Dynamical modeling of large scale asymmetries in the \bp\ dust disk.
}
\titlerunning{Dynamical modeling of large scale asymmetries in the
\bp\ dust disk}
\author{J.C. Augereau\inst{1} \and R.P. Nelson\inst{2} \and
A.M. Lagrange\inst{1} \and J.C.B. Papaloizou\inst{2} \and
D. Mouillet\inst{1}
}
\authorrunning{J.C. Augereau et al.}
\offprints{J.C. Augereau}
\institute{ Laboratoire
 d'Astrophysique de l'Observatoire de Grenoble, Universit\'e J.
 Fourier / CNRS, B.P. 53, F-38041 Grenoble Cedex 9, France \and
 Astronomy Unit, School of Mathematical Sciences, Queen Mary
 \& Westfield College, Mile End Road, London E1 4NS, UK}
\mail{augereau@obs.ujf-grenoble.fr} 
\date{Received 27 october 2000 / Accepted 19 january 2001}
\abstract{We report a new and complete model of the \bp\ disk,
which succeeds in accounting for both the surface brightness
distribution, warp characteristics, the outer ``butterfly'' asymmetry
as observed by HST/STIS in scattered light, as well as the infrared
emission.
Our model includes the presence of a disk of planetesimals extending
out to 120--150\,AU, perturbed gravitationally by a giant planet on an
inclined orbit, following the approach of \citet{mou97b}. At any time,
the planetesimal disk is assumed to be the source of a distribution of
grains produced through collisional evolution, with the same initial
orbital parameter distribution. The steady state spatial grain
distribution is found incorporating the effects of radiation pressure
which can cause the distribution of the smallest particles to become
very distended.
With realistic assumptions about the grains' chemical properties, the
modeling confirms the previously evident need for an additional
population of hot grains close to the star, to account for the
12\,$\mu$m fluxes at short distances from the star. It also indicates
that this population cannot explain the outer 12\,$\mu$m flux
distribution when the effects of gravity and radiation pressure
determine the distribution.  Very small grains, produced by
collisions among aggregates, are tentatively proposed to account for
this 12\,$\mu$m outer emission. \keywords{Stars: \cs\ matter -- Stars:
\bp}}
\maketitle
%
%
%
\section{Introduction}
The \bp\ gaseous and dusty disk has been extensively studied for 15
years. The dust disk, seen nearly edge-on, extends to at least a
distance of 100\,AU from the central star, with a sharp decrease in
the surface brightness distribution of scattered light beyond about
120\,AU. The dynamics of the small grains producing the scattered
light is determined by radiation pressure, collisions with other
grains and/or evaporation and to a lesser extent, Poynting-Robertson
drag. The images at 12\,$\mu$m show that the maximum of the dust
surface density distribution is located between 80\,AU and 100\,AU
\citep[][assuming the Hipparcos distance of 19.28\,pc from \citet{cri97}
for \bp]{pan97}. The total {\it dust} mass, as measured from
sub-millimeter data, ranges between a few and a few tens of lunar
masses \citep{zuc93,hol98,den00}.

The 10\,$\mu$m silicate spectrum \citep{kna93} appears to mimic that
of Halley's comet but not that of comet Hale-Bopp \citep{lag99}. The
10\,$\mu$m features are best fitted by the emission of small
(submicronic) crystalline silicates \citep{kna93,li98}. It was
suggested that these grains could result from comet evaporation, a
scenario that had already been proposed to explain the very peculiar
spectroscopic activity \citep{aml87,fer87}. Furthermore, \citet{lec96}
proposed that the appearance of the entire disk, as seen in scattered
light, could be explained by grains released from evaporating comets
at distances of 15--30\,AU and subsequently pushed out by radiation
pressure.
\subsection{Asymmetries in the dust disk}
The Northeast and Southwest extensions of the dust disk have been
found to be asymmetric in scattered light as well as in thermal
emission. At visual wavelength, \citet{kal95} noticed five radial or
vertical asymmetries beyond 150\,AU. In the Northeast extension for
example, the vertical brightness distribution in scattered light is
asymmetric with respect to the disk mid-plane out to hundreds of AU
and this situation is reversed in the Southwest extension
(``butterfly'' asymmetry). \citet{kal00} and \citet{lar01} recently
proposed that this asymmetry is produced by a close encounter with a
M0V star in the last $10^5$ years.

In the inner part of the disk (at about 60-70\,AU) the 12\,$\mu$m
images reveal a factor of 3 side to side brightness asymmetry
\citep{pan97} which is much smaller at optical and near-infrared
wavelengths \citep[][at most a factor of 1.5]{mou97a}. Orbiting
evaporating bodies on eccentric orbits with the same longitude of
periastron can produce such axial asymmetries
\citep{lec96,lec98}.

Adaptive optics \citep{mou97b} and HST observations
\citep{bur95,hea00} revealed the inner warping of its dust disk at
about 70\,AU and in the same direction as the ``butterfly'' asymmetry.
\citet{mou97b} modeled this warp as being due to grains
which followed the distribution of a parent body planetesimal disk
which was perturbed by the gravitational influence of a Jupiter-like
planet on an orbit inclined at 3 degrees to the disk mid-plane.
\subsection{Propagation of the warp}
The important effect of perturbation by the giant planet is to cause
the parent body disk to precess differentially. The more rapid
precession in the inner regions causes the parent body disk to become
coplanar with the planetary orbit.  At large distances the inclination
of the parent body disk retains its initial value.  The warp is
located in the transition region where the local orbital precession
timescale is comparable to the age of the system.  The location of the
warp accordingly propagates outwards with time.

\citet{mou97b} found an approximate relationship between the position
of the warp, the mass ratio of planet to star, its orbital (assumed
circular) radius and the age of the system in the form:
\begin{eqnarray*}
\log\left(\frac{R_w}{10{\rm AU}}\right)=
0.29\,\log\left(\frac{M}{M_*}\left(\frac{D}{10{\rm AU}}\right)^2
\frac{t}{t\dma{unit}}\right)-0.2 \,\,.
\label{warppos}
\end{eqnarray*}
Here $M_*, M, R_w, D$, $t$ and $t\dma{unit}$ are the mass of the star,
the mass of the planet, the radius of the warp {\it in the parent body
disk}, the planetary orbital radius, the age of the system and the
time unit $\sqrt{(10{\rm AU})^3/(GM_*)} \sim 5.2\,y$ respectively. For
example, to obtain $R_w =70 {\rm AU},$ $ M/M_* =10^{-3}$ and
$D=10$\,AU requires the age of the system to be $\sim 2.1\times
10^7\,y.$ Such orbital parameters for the proposed perturbing planet
allow the spectroscopic activity to be successfully reproduced
\citep[][and references therein]{beu00}.

Note that when, as assumed here, the warp location is significantly
larger than the orbital radius, that location depends only on the
product of the age of the system, the square of the orbital radius and
the planetary mass ratio.  We comment that, because it does not
incorporate a non axisymmetric perturbation, the model proposed by
\citet{lec96} cannot account for the inner warp, but the influence of
radiation pressure, clearly revealed by their study, is certainly of
great importance.
\subsection{Aims of the present modeling}
Complete modeling of the disk has not been performed so far. In this
paper we formulate a model of the disk which reproduces both the
scattered light images, in particular the warped disk and the
``butterfly'' asymmetry, and also the infrared (IR) images and fluxes.
In Section \ref{model}, we outline the basis of this model and Section
\ref{scatt} presents the scattered light images which can be compared
with the recent HST/STIS data \citep{hea00}.  In section \ref{therm},
we synthesize the IR 12\,$\mu$m image and compare with the
observations.  In section \ref{discuss} we summarize and discuss the
proposed model in a more general context.
%
\section{The basic model}
\label{model}
\subsection{The parent body disk}
The short lifetime of the grains, in comparison to the age of the
system, as a result of the action of collisions and radiation pressure
leads to the hypothesis of the existence of a parent body disk
consisting of planetesimals with sizes larger than 1\,$m$ which acts
as a source for the grains as collision products \citep[see for
example the discussion in][and references therein]{mou97b}.

As in \citet{mou97b} we suppose that the physical collision time in
the parent body disk is sufficiently long compared to the age of the
system that it can be approximated as a collisionless disk lying
between $1.5D$ and $15D.$ The disk aspect ratio $H/r$ is taken to be
0.1. Through our modeling procedure we are able to construct arbitrary
surface density profiles. This is done by constructing parent body
disks consisting of weighted contributions from 100 elementary rings
composed of $10^4$ particles and with a ratio of inner and outer radii
equal to $10^{1/99}\sim 1.0235$. This is possible because the
particles are collisionless.
\subsection{Orbital parameters of the perturbing planet}
For working parameters we adopt $D= 10{\rm AU},$ being also the model
unit of distance, planet to star mass ratio $ M/M_* = 10^{-3},$
planetary orbital eccentricity of zero with an inclination of 3$\degr$
with respect to the initial mid-plane of the parent body disk.
However, as above, we note that the important parameter is $MD^2$
rather than $M$ and $D$ separately. \citet{mou97b} also show that the
propagation of the warp is unaffected for eccentricities of the
planetary orbit up to 0.5. however, large eccentricities may lead to
side to side (radial) asymmetries of the parent body disc which in
turn may lead to some asymmetries in the generation of small dust
particles. The study of asymmetries produced in this way is beyond the
scope of this paper.
\subsection{ Parent body disk precession}
The parent body disks are modeled here using $10^6$ particles. It is
not possible to carry out orbital integrations over the system
lifetime for such a number of particles.  Instead, after realizing the
parent body disk by allocating particles a random inclination and
eccentricity consistent with the aspect ratio, the system was allowed
to phase mix by integrating for 20 outer orbital periods. At this
stage the orbital plane of each particle was precessed at the local
rate for the age of the system while each particle retained its
orbital phase.  Finally the system was phase mixed again.  This
procedure is equivalent to evolving the orbits using the time averaged
potential of the planet and its validity has been checked by
comparison with complete integrations over time spans where these were
feasible.  The local precession rate for near circular orbits was
taken to be \citep{mou97b}:
\begin{eqnarray*}
|\omega_p| ={3GM D^2\over 4 \Omega r^5}\,\,\,.
\end{eqnarray*}
Here $r$ is the orbital radius and $\Omega= \sqrt{GM_*/r^3}$ is the
local Keplerian frequency.  The age of the system adopted here,
corresponding to the mass ratio $M/M_* =10^{-3},$ was $2.6\times 10^7
y$ consistent with the recent estimated age of \bp\ from \citet{bar99}
and which yields $R_w\simeq 75\,$AU.  But note as above that the
significant parameter is the product of the age and the mass ratio.

\begin{figure}[tbph]
\includegraphics[angle=0,width=0.49\textwidth,origin=br]{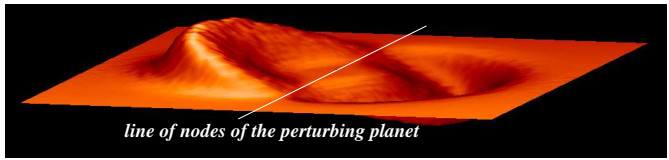}
\caption[]{Vertical position of the mean PB disk mass assuming
  the surface density distribution of Figure \ref{density}. The
  vertical and radial scales are not the same in this Figure. The
  warped disk extends out  to 140--150\,AU with a maximum close to
  75\,AU from the star and a vertical extension of $\sim$4\,AU to the
  initial mid-plane of the PB disk.}
\label{PBwarp}
\end{figure}

The configuration of the warped parent body disk, obtained using the
above procedures, with the surface density we adopted is illustrated
in Figure \ref{PBwarp}.
\subsection{Simulation of the dust disk}
We assume that each planetesimal (parent body or hereafter PB) is the
progenitor of small particles (produced for instance by collisions)
that are significantly affected by the radiation pressure of the
central star.  We assume no initial relative velocity between the PB
and the smallest particles produced.  A consequence of this assumption
and the near circular motion of the PB is that the periastron of a
dust particle equals that of the PB. The effect of radiation pressure
is quantified by the ratio of the force due to radiation pressure to
that due to the gravity of the central star, $\beta\dma{pr} =
F\dma{rad}/F\dma{grav}.$ The case $\beta\dma{pr} = 0$ corresponds to a
bound dust particle with the same motion as its PB. When
$\beta\dma{pr} = 0.5$, particles have zero energy and so are just
unbound. We consider 11 values of $\beta\dma{pr}$ equally spaced
between and including 0.45 and 0.001.  For each value of
$\beta\dma{pr}$, a dust distribution is generated from the PB by phase
mixing $10^{6}$ particles for 20 orbital periods measured
at the outer radius of the disk.  The time-scales
required for phase mixing are always short compared to the time-scale
for propagating the warp.  Also the optical depth of the disk is large
enough that the timescale for destruction of the dust by collisions is
shorter than any drag timescale due to the Poynting-Robertson effect
\citep[see also][]{mou97b}. The Poynting-Robertson effect then is
neglected here.
%
\subsection{Modeling of the scattered light and thermal images}
\subsubsection{Efficient modeling of the PB surface density distribution}
To construct the final image of the disk in scattered light or in
thermal emission arising from any PB surface density distribution in
practice, a radial weight function is applied to the contribution from
each dust particle depending on the elementary ring between $1.5D$ and
$15D$ from which it originated (the distance to this ring also
corresponds to the periastron distance of the dust particle).  Several
$\Sigma(r)$ distributions can therefore be investigated using this
technique in a reasonable amount of time.
\subsubsection{Grain properties}
The image of the disk also depends on the physical properties of the
grains.  Here we have adopted the comet dust model proposed by
\citet[][and references therein]{li98} to reproduce the \bp\ spectral
energy distribution (SED).  In this model, the dust consists of porous
aggregates made from a silicate core coated with organic refractories
\citep[see also][]{pan97}. A more detailed account of the grain model
used here is given in \citet{aug99}.  Initially, we fix the porosity
of the grains to be $0.95$ and assume that most are amorphous
\citep{li98,lag99}.
\subsubsection{Presence of ice}
Water ice may be present on the surfaces of the elementary submicronic
particles that compose an aggregate \citep[][and references
therein]{gre90}.  The amount of ice is quantified by a volume
percentage in addition to that of vacuum (due to the porosity). A
grain is assumed icy if it is produced further than the ice
sublimation distance. That distance depends on the grain size and
amount of ice (typically between 20\,AU for grains larger than
$\sim100\,\mu$m and 100\,AU for grains smaller than $\sim1\,\mu$m).
Due to the effect of radiation pressure, non-icy grains can be present
at very large distances as well as icy grains, possibly mixing two
dust chemical compositions at a given distance from the star.
\subsubsection{Grain sizes}
For the range of $\beta\dma{pr}$ values we consider, there is a
one-to-one relationship between the value of $\beta\dma{pr}$ and the
grain size $a$ which can be expressed in the form
$a=K\beta\dma{pr}^{-1}$ \citep[][for instance]{art88}, where $K$
depends on the optical properties of the grain which are computed
using the Mie theory \footnote{The Maxwell-Garnett effective medium
  theory is used to compute the complex index of refraction of an
  aggregate.}  \citep{boh83}. Some examples of $K$ values for
different amounts of ice and porosities can be derived from the values
of $a\dma{\beta\dma{pr}=0.5}$ given in Table \ref{IRAS}. A grain size
distribution with $ dn(a)/da \propto a^{-3.5}$ is assumed to take into
account the relative number $n(a)$ of grains with radius greater than
$a$ produced in the disk by collision processes. For each
$\beta\dma{pr}$ value (or grain size), the appropriate relative number
density and thus the contribution to a scattered light or thermal
emission image is computed. The 11 independent images then are
correctly superimposed with appropriate weight factors so as to
produce a final image of the disk for any required PB surface density
distribution or wavelength of interest.
\subsubsection{Orientation of the disk to the line of sight}
In order to complete the above procedure, the disk orientation with
respect to the observer has to be specified. This introduces two
additional parameters: the line of nodes of the perturbing planet and
the disk inclination with respect to a plane normal to the line of
sight.
%
\section{Scattered light images}
\label{scatt}
\subsection{PB surface density distribution}
The first stage in the modeling procedure is to find a surface density
distribution for the PB disk that leads to consistent results with
both scattered light observations and thermal emission data. The range
of possible surface density distributions can be restricted using the
following observational constraints:\h
1/ Both the resolved 12\,$\mu$m images \citep{pan97} and SED modeling
\citep{li98} indicate that little dust lies inside
$\sim$50\,AU.\h
2/ Beyond about 120\,AU, the radial surface brightness profile in
scattered light revealed by STIS is roughly proportional to $r^{-5}$
\citep[][]{hea00}. For an edge-on disk like the \bp\ one, this is in
very good agreement with the profile expected from small dust
particles produced by colliding large bodies confined within 120\,AU
and blown out by radiation pressure \citep{lec96}.\h
Because dust particles trace the distribution of the PB disk
\citep{mou97a}, qualitatively, these remarks imply that most of the PB
disk mass is contained within an annulus lying between $\sim$50\,AU
and 120\,AU.
\subsection{Surface brightness distribution}
\begin{figure}[tbph]
\includegraphics[angle=90,width=0.49\textwidth,origin=br]{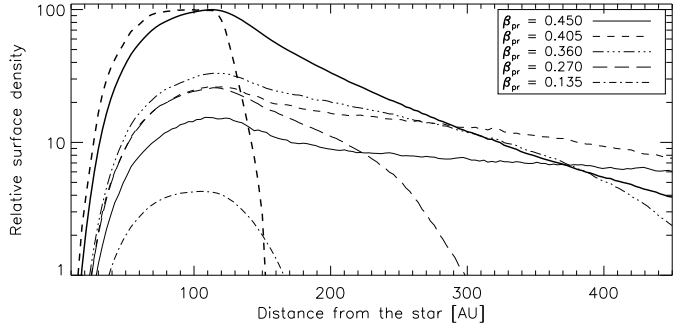}
\caption[]{Surface density of the PB disk (bold-dashed line) that gives
  a good fit to the observed surface brightness distribution in
  scattered light. The resulting dust surface density integrated over
  the grain size (or $\beta\dma{pr}$) distribution is displayed  with a
  bold-solid line.  Both profiles are normalized to 100 at their
  maximum.  The relative grain surface densities (thin lines) for five
  values of $\beta\dma{pr}$ ranging between 0.135 and 0.45 are also
  superimposed for comparison.}
\label{density}
\end{figure}
\begin{figure}[tbph]
\includegraphics[angle=90,width=0.49\textwidth,origin=br]{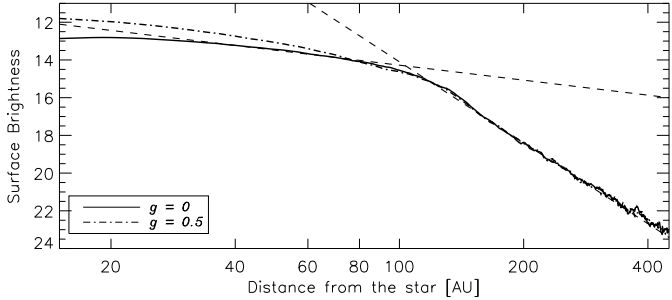}
\caption[]{Surface brightness distributions in mag.arcsec$^{-2}$ along
  the mid-plane of the disk in scattered light (R band) assuming the PB
  surface density shown in Figure \ref{density}. The  solid line
  assumes isotropic scattering properties whereas the dash-dotted line
  assumes a \citet{hen41} phase function with an asymmetry factor
  $|g|$ of $0.5$. The two dashed lines represent radial power laws
  with indices -1.1 and -5.5 as measured on the HST/STIS images.}
\label{scBDS}
\end{figure}

Applying the observational constraints described above, we tried
several PB body surface density distributions. We were able to find a
solution (Figure \ref{density}) that gives a fit to the scattered
light images (Figure \ref{scBDS}).

In this solution, the distribution of PB is confined inside
$\sim$150\,AU. Due to the radiation pressure acting on the grains, the
outer part of the disk is filled  with small particles. This can be seen
in Figure \ref{density}, where we have plotted the resulting surface
density distributions for grains of different sizes. The final surface
density of the grains averaged over the grain size distribution peaks
at about 110--120\,AU from the star, close to the assumed outer edge
of the parent disk (Figure \ref{density}).

At distances greater than about 30\,AU, the main shape of the mid-plane
surface brightness distribution in scattered light from \citet{hea00}
is well matched depending somewhat on the precise anisotropic
scattering properties (Figure \ref{scBDS}).

A radial power law index between $-5$ and $-5.5$ is measured at
distances larger than 120\,AU when the dust disk is seen almost
edge-on as predicted by \citet{lec96} and indeed observed with the
HST/STIS instrument.

\begin{figure}[tbph]
\includegraphics[angle=90,width=0.49\textwidth,origin=br]{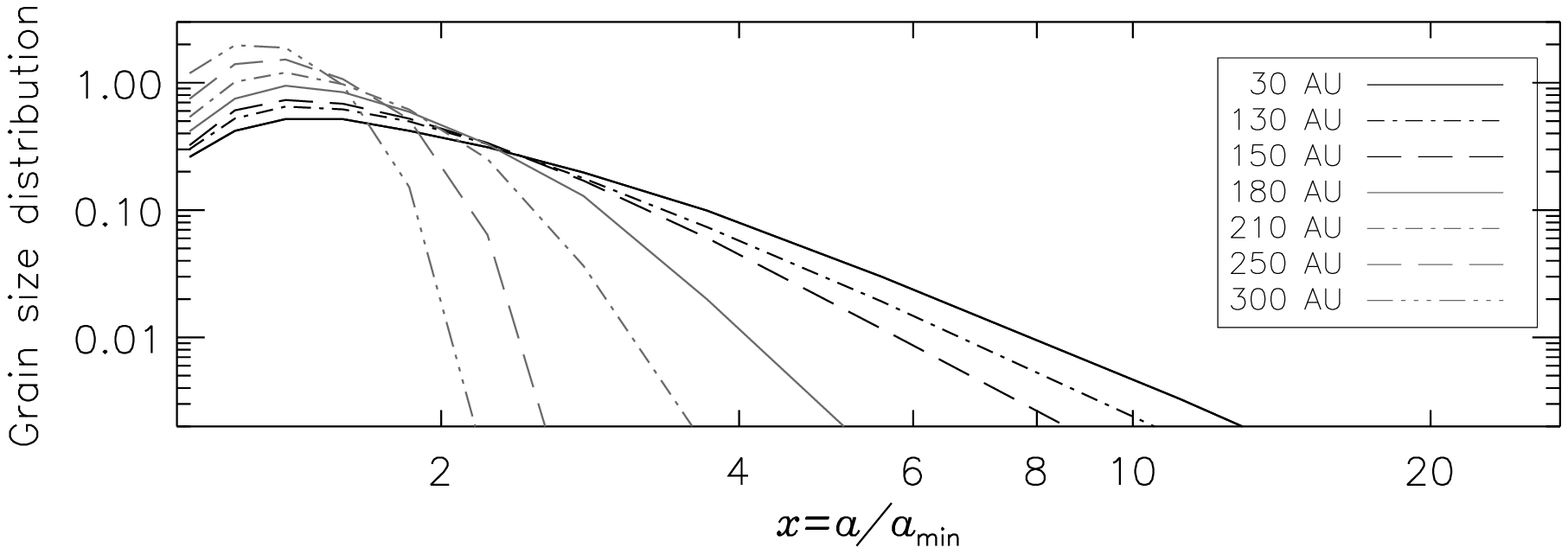}
\h
\includegraphics[angle=90,width=0.49\textwidth,origin=br]{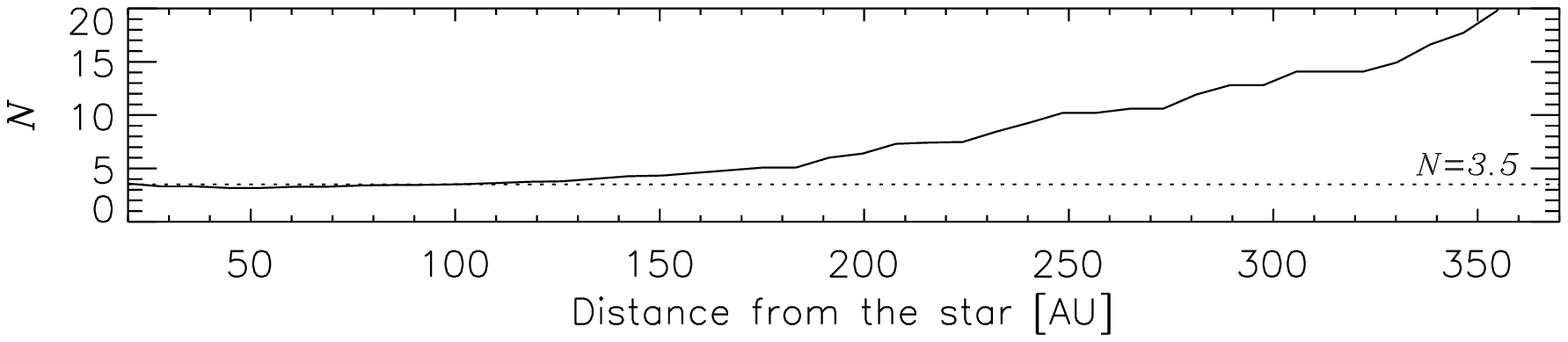}
\caption[]{{\it Upper panel:} Grain size distribution  at different
  distances from the star.  In each case the distribution is
  normalized such that the integral over $a$ has been normalized to
  $1$. {\it Lower panel:} $N$ index assuming that the grain size
  distribution $|dn(a)/da|$ is proportional to: $(1-x^{-1})^M\times
  x^{-N}$ with $x=a/a\dma{min}$ and
  $a\dma{min}=a_{\beta\dma{pr}=0.5}=2K$ (see Table \ref{IRAS}). In
  this example, $M$ is between $1.4$ and $2$ inside $\sim 150$\,AU and
  up to $4$ at larger distances.}
\label{sizedist}
\end{figure}

Importantly, due to the effect of radiation pressure, the distribution
of the grain sizes is a function of the distance from the star as
shown in Figures \ref{density} and \ref{sizedist}.
\subsection{Small  scale vertical asymmetry (the warp)}
\begin{figure*}[tph]
\includegraphics[angle=90,width=0.99\textwidth,origin=br]{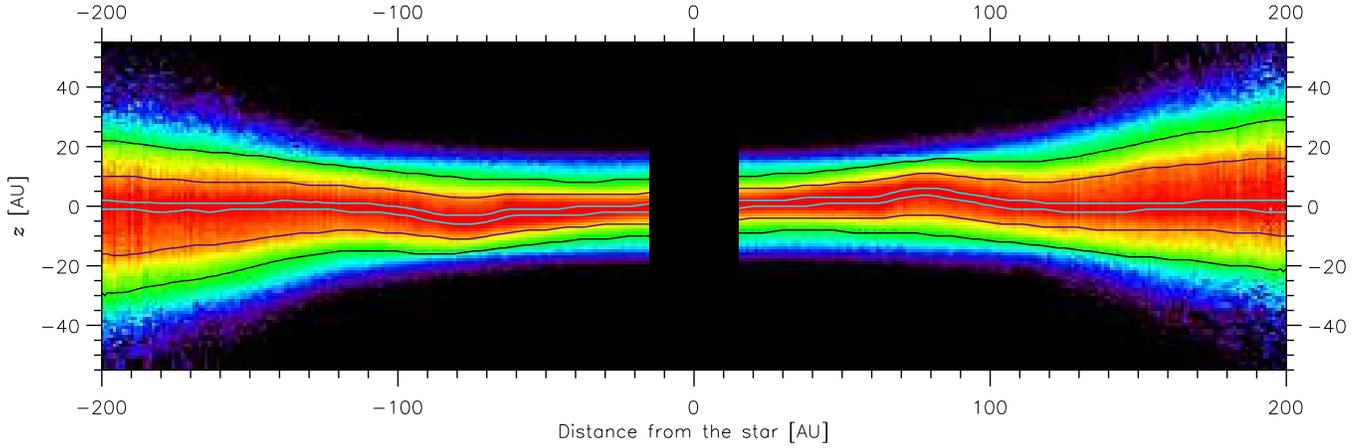}
\caption[]{The warped disk normalized for  each vertical cut to its
  maximum surface brightness. To compare with Figures 10 from
  \citet{hea00} and to emphasize the vertical asymmetry, contours at
  99\%, 50\% and 10\% of the maximum surface brightness have been
  superimposed.}
\label{diskNORM}
\end{figure*}
Due to the effect of the assumed perturbing planet, the precession of
the orbital planes of PB within 70--80\,AU results in a distribution
with average inclination coincident with the planetary orbit. Beyond
70--80\,AU and up to the assumed outer edge of the PB disk (close to
140--150\,AU), the PB orbits have undergone only partial precession.
Therefore, the mean inclination of their orbital planes to that of the
initial disk ranges between 3$\degr$ and about 0.5$\degr$.

\citet{mou97b} showed that the warp, or the short scale asymmetry of
the vertical position of the maximum surface brightness, can be
reproduced assuming all particles followed the distribution of the
parent body disk. Adding the effects of radiation pressure acting on
grains with a given size distribution to the model of \citet{mou97b}
causes a modest shift of the warp position, and so does not affect the
conclusions on the propagation and observability of the warp as seen
in Figure \ref{diskNORM}, which can be directly compared to the
HST/STIS images \citep{hea00}.
\begin{figure}[tbph]
\includegraphics[angle=90,width=0.49\textwidth,origin=br]{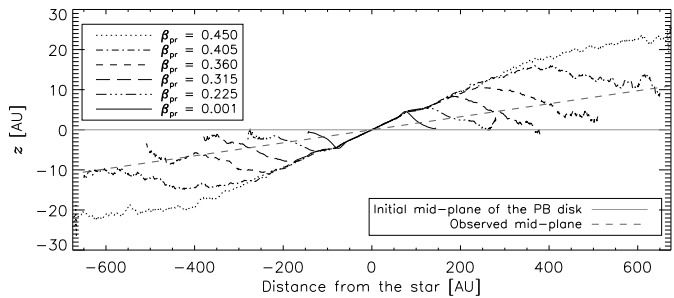}
\caption[]{The location of the maximum surface brightness
  for a vertical cut as a function of distance, for disks composed of
  grains with a single size as indicated, seen in scattered light. The
  line for $\beta\dma{pr}=0.001$ traces the PB disk. The line for
  $\beta\dma{pr}=0.45$ indicates the way the small scale asymmetry
  (warp) can be extended outwards as a result of radiation pressure.
  The position of the observed mid-plane to the original mid-plane of
  the PB disk is superimposed in this particular case (line of nodes
  of the planet aligned with the line of sight and disk seen
  edge-on).}
\label{warp}
\end{figure}
\subsection{The large scale vertical (``butterfly'') asymmetry}
\begin{figure*}[tbph]
  \includegraphics[angle=0,width=0.99\textwidth,origin=br]{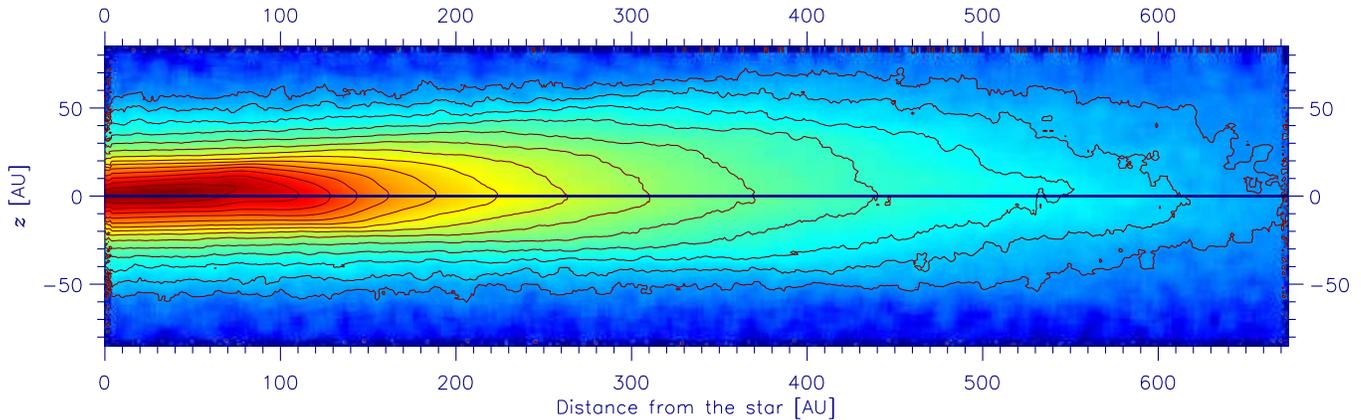}
\caption[]{Modeled disk at large distances from the star. A large scale
  asymmetry with respect to the mid-plane is found, similar to the
  observed ``butterfly'' asymmetry.}
\label{wing}
\end{figure*}
We now investigate whether our model can account for the ``butterfly''
asymmetry. Basically, radiation pressure greatly extends the surface
density distribution of the smallest particles, {\it i.e.} those which
have the larger $\beta\dma{pr}$ values, beyond that of the PB disk
(figure \ref{density}).  Accordingly, the asymmetry in the vertical
location of the surface brightness maximum originating in the PB disk
can be transmitted outwards as illustrated in Figure \ref{warp}. The
result is a flared-like disk, with the so-called butterfly asymmetry.

For illustrative purpose, we focus on the particular case for which
the line of nodes of the planetary orbit and the line of sight are
aligned \citep[the observability of the warp as a function of the
location of the line of nodes is discussed in][]{mou97b}. The
simulated disk in scattered light then convincingly exhibits a large
scale asymmetry as observed in STIS data up to 200\,AU (Figure
\ref{diskNORM}).

\begin{figure}[tbph]
\begin{center}
  \includegraphics[angle=90,width=0.45\textwidth,origin=br]{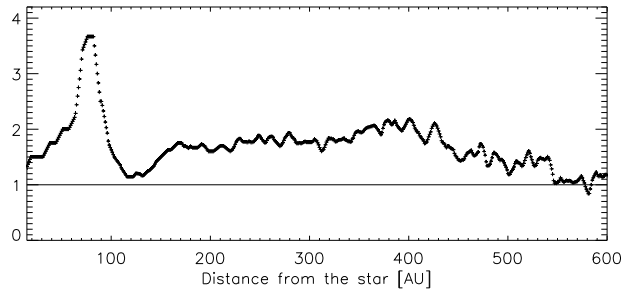}
\caption[]{Ratio of half widths at half maximum above and below the
  observed mid-plane for a vertical cut as a function of distance and
  for the extension of the disk seen in Figure \ref{wing}.}
\label{asym}
\end{center}
\end{figure}

It is worth noticing that this asymmetry is present much further out
than 200\,AU (Figure \ref{wing}), in agreement with the large scale
images of \citet{kal95} for instance. Figure \ref{asym} quantifies the
radial dependence of the asymmetry by the ratio of half widths at half
maximum above and below the observed mid-plane. The ratio evidences
the position of the warp at 70--80\,AU and shows that the disk remains
asymmetric up to at least 500\,AU as observed. In other words, the
asymmetric distribution of PB responsible for the observed warp at
70--80\,AU, and assumed to result from the perturbation of an inclined
planet, leads naturally to a large scale asymmetry very similar to the
``butterfly'' asymmetry. Therefore, the observability of the
``butterfly'' asymmetry depends as well on the inclination of the
planet to the PB disk mid-plane (3$\degr$ in this model), and on the
line of nodes of the planet to the line of sight.
\subsection{Summary}
The dynamical approach we have followed allows us to reproduce in a
consistent way most observations in scattered light, namely: the
surface brightness distribution, the small scale (warp) and large
scale (``butterfly'') asymmetries, assuming a planetary perturber and
taking into account the effects of radiation pressure acting on the
grains. But as expected, the model does not explain the NE--SW
asymmetries.
\section{Thermal emission data}
\label{therm}
\subsection{IRAS fluxes}
\begin{table*}[t] 
\begin{center}
\caption{\label{IRAS} Thermal emission in IRAS bands for different dust
  models. According to \citet{li98}, dust models with $P=0.98$ and
  $P=0.95$ assuming icy grains give a satisfactory fit to the overall
  SED. The percentages of ice have been adjusted to respect the ice to
  silicate core-organic refractory mantle mass ratio of 1 from
  \citet{li98}. Dust grains with $P=0.5$ have a mass ratio of 0.25.
  Dust models with non-icy grains are also given for comparison.  In
  each case, the dust surface density has been normalized in order to
  reproduce the scattered light profile and then the integrated
  thermal emission computed.  With this procedure, the computed IR
  fluxes then also depend indirectly on the grain anisotropic
  scattering properties. There, we assume isotropic scattering
  properties, but the fluxes are increased by a factor of about 1.4 if
  the asymmetry factor $|g|$ increases to 0.5 and the ratio of the
  observed to computed flux is accordingly decreased by the same
  amount.}
\begin{tabular}[t]{cccccccc}
\hline
 & IRAS & $\begin{array}{c} P=0.98 \\ 0\textrm{\% ice} \end{array}$ & $\begin{array}{c} P=0.98 \\ 4\textrm{\% ice} \end{array}$ & $\begin{array}{c} P=0.95 \\ 0\textrm{\% ice} \end{array}$ & $\begin{array}{c} P=0.95 \\ 10\textrm{\% ice} \end{array}$ & $\begin{array}{c} P=0.5 \\ 0\textrm{\% ice} \end{array}$ & $\begin{array}{c} P=0.5 \\ 50\textrm{\% ice} \end{array}$ \\ 
\hline
Flux at 12\,$\mu$m [Jy] & 1.64$\pm$0.1 & 0.040 & 0.056 & 0.037 & 0.047 & 0.017 & 0.017 \\
Flux at 25\,$\mu$m [Jy] & 10.1$\pm$0.5 & 3.5 & 2.1 & 3.4 & 1.9 & 2.4 & 1.6 \\
Flux at 60\,$\mu$m [Jy] & 18.8$\pm$1 & 12.5 & 16.3 & 12.4 & 16.3 & 11.9 & 13.6 \\
Flux at 100\,$\mu$m [Jy] & 11.2$\pm$1 & 10.4 & 10.2 & 10.3 & 10.2 & 9.8 & 9.7 \\
\hline
$a_{\beta\dma{pr}=0.5}=2K$ [$\mu$m] $\left\{\begin{array}{l} \textrm{no ice} \\ \textrm{ice} \end{array}\right.$&  & $\begin{array}{c} 92.7 \\ \, \end{array}$ & $\begin{array}{c} 92.7 \\ 46.5 \end{array}$ & $\begin{array}{c} 37.1 \\ \, \end{array}$ & $\begin{array}{c} 37.1 \\ 18.9 \end{array}$ & $\begin{array}{c} 3.7 \\ \, \end{array}$ & $\begin{array}{c} 3.7 \\ 3.0 \end{array}$ \\
\hline
$\begin{array}{c} \textrm{Ratio of observed to} \\
\textrm{computed flux at 12\,$\mu$m} \\
 \textrm{further than $\sim$30\,AU (NE side)} \end{array}$ & & 6 & 12 & 7 & 15 & 16 & 26\\
\hline
\end{tabular}
\end{center}
\end{table*}
A fully consistent disk model should reproduce both the scattered
light images and thermal data. Our aim is to estimate the thermal IR
(12, 25, 60 and 100\,$\mu$m) fluxes produced by the distribution found
in Section \ref{scatt}. Table \ref{IRAS} summarizes the computed
fluxes assuming the dust model proposed by \citet{li98} for different
porosities and amounts of ice.

The 100\,$\mu$m and 60\,$\mu$m emissions (the latter slightly
depending on the amount of ice) are correctly matched by the dust ring
made of grains at the blow-out size limit that reproduces the
scattered light profile. Such a consistency strongly supports our
dynamical approach.

On the other hand, the simulated 12\,$\mu$m and 25\,$\mu$m fluxes are
smaller than observed. This shows that an additional population of
hotter grains is required, in total agreement with previous works
showing that most of the 12\,$\mu$m emission comes from a region close
to the star \citep[see for example][]{kna93}, which is also confirmed
by the resolved images from \citet{pan97}. This close population also
certainly contributes to the integrated 25\,$\mu$m flux.
\begin{figure}[tbph]
\includegraphics[angle=90,width=0.49\textwidth,origin=br]{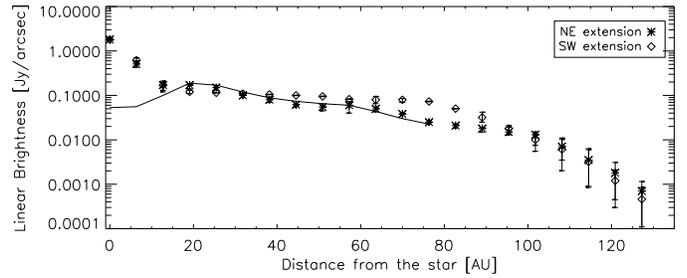}
\caption[]{Linear brightness distribution in thermal emission at
  $\lambda=12\,\mu$m assuming the PB surface density distribution
  shown in Figure \ref{density} and non-icy grains larger than $2K$
  with $P=0.95$.  At each distance from the star, this linear
  brightness has been computed by multiplying the maximum of flux
  along the vertical axis by the vertical thickness (in arcsec) of the
  disk at $12\,\mu$m given by the Figure 4 from \citet{pan97} and
  available for distances smaller than 70\,AU. To compare with the
  \citet{pan97} resolved emission, the predicted fluxes have also been
  multiplied by a factor of 7, indicating that most of the grains that
  contribute to the $12\,\mu$m emission at distances larger than
  20--30\,AU are not taken into account.}
\label{thBDSisotrope}
\end{figure}
\subsection{ The 12\,$\mu$m resolved images}
Most of the 12\,$\mu$m flux originates within $\sim$20\,AU from the
star as seen before. However, as shown by \citet{pan97}, a small part
of the 12\,$\mu$m flux comes from regions up to $\sim$100\,AU. Our aim
is here to see whether our model accounts for this ``outer''
12\,$\mu$m emission.

Simulated and observed 12\,$\mu$m images compare well in terms of
radial shape down to about 20--30\,AU (and especially when the most
porous grains are not too icy, Figure \ref{thBDSisotrope}) but not in
terms of flux. The ratio of observed to computed flux is given in the
last line of the Table \ref{IRAS} for different dust properties. This
shows that an additional population of efficient 12\,$\mu$m emitters,
and at  the same time bad scatterers at near IR wavelengths, is needed
between 20--30\,AU and 100\,AU. Very small, typically submicronic,
grains are a priori good candidates. For example, the
emission/absorption at 12\,$\mu$m to scattering at 1\,$\mu$m
efficiency ratio is a few hundred for $0.1\,\mu$m grains with porosity
$P=0.95$ and 3 orders of magnitude larger if the grains are only
$10^{-2}\,\mu$m in size.
%
\subsection{Possible scenarios to account for the 12\,$\mu$m missing
 flux at large distances}
Core-mantle grains smaller than $a_{\beta\dma{pr}=0.5}=2K$ have
$\beta\dma{pr}$ ratios larger than 0.5. A close inspection of 
Figure \ref{density} indicates that particles with $\beta\dma{pr}$
ratios larger than typically 0.4 have quite similar surface densities
and tend to mimic that of the $\beta\dma{pr}=0.45$ grains which are
almost just unbound.  In the following, we will use this property to
simulate a dust disk with very small particles.

We propose hereafter possible scenarios {\it a priori} able to explain
the presence of very small grains at large distances, and we check
whether or not they may quantitatively reproduce the 12\,$\mu$m flux
at large distances under reasonable assumptions.
\subsubsection{First scenario: Small grains produced by cometary
  evaporation}
A first possibility is that the submicronic grains produced close to
the star, for instance by comet evaporation, responsible for most of
the integrated 12\,$\mu$m flux and required to explain the 10\,$\mu$m
silicate spectroscopic features, also supply the disk beyond
$\sim$30\,AU with enough small dust particles through radiation
pressure. This would give an attractive solution to the lack of
12\,$\mu$m emission at distances larger than $\sim$30\,AU in our
model.

We therefore assumed a population of very small grains close to the
star and let them evolve under radiation pressure. It appears that,
whatever the assumptions on the surface density distribution of the
inner population, the radial shape of the 12\,$\mu$m flux distribution
does not match the one found by \citet{pan97} beyond $\sim$30\,AU.
This is mainly due to the radial decrease of the grain temperature
with distance from the star which tends to result in most of the
emission at 12 \,$\mu$m
 coming from close to the star leading either to
very to small contributions further than $\sim$30\,AU or to radial
profiles for the emission not consistent with the resolved observations.

A population of submicronic grains certainly exists close to the star
but our modeling indicates that additional grains produced further
than $\sim$30\,AU should be present to account for the radial shape of
the 12\,$\mu$m emission.
\subsubsection{Second scenario: Small unbound grains below the blow-out limit}
We will now consider the dust disk that reproduces the scattered light
data and extrapolate the initial $a^{-3.5}$ grain size distribution to
grains with $\beta\dma{pr}$ ratios larger than 0.5. We consider
non-icy grains for simplicity and $P=0.95$ which gives good fits to
the large wavelength fluxes.

In other terms, we simply extend the grain size distribution that
successfully fitted the scattered light images in Section \ref{scatt}
to a minimum grain size of about 0.1\,$\mu$m instead of a few tens of
micrometers. This actually allows  us to fit the 12\,$\mu$m emission
further than $\sim$20\,AU. The 60 \,$\mu$m
 and 100\,$\mu$m integrated fluxes
are unchanged as well as the scattered light profile inside 120\,AU
because the amount of large grains is not affected by the reduction in
minimum grain size. However, beyond 120\,AU, the scattered light
surface brightness profile changes: it follows now a $r^{-3.5}$ radial
power law (instead of $r^{-5}$). $r^{-3.5}$ is not consistent with
STIS data, but is in better agreement with previous data
\citep[e.g.][]{gol93,kal95,mou97b}.

It is therefore premature to conclude whether or not this
extrapolation is correct on the sole basis of the surface brightness
distribution at large distances. Moreover, the rate of production of
such small grains must be continuous and more efficient than indicated
by a simple extrapolation from the larger grains, since being unbound,
their life-time in the disk is very short, in fact $1$ to $10^3$ times
shorter than grains with sizes just above the blow out limit. The mass
loss rate from the system due to the blow out of these grains is also
correspondingly larger.
\subsubsection{Third scenario: Small compact bound  grains produced by collisions}
The physics of aggregates, present in \cs\ environments, is complex and
not well understood. Catastrophic disruption of an aggregate can occur
if the impact velocity is high enough \citep[typically a few tenths of
km.s$^{-1}$,][for instance]{dom97}, producing nearly compact
submicronic grains.
In the vicinity of \bp, such grains smaller than $\sim 10^{-2}\,\mu$m
may have $\beta\dma{pr}$ ratios smaller than 0.5 depending on their
precise chemical composition, {\it i.e.} if they are not too
refractory \citep[e.g.][]{art88}. This leads ultimately to a
discontinuity in the grain size distribution which could appear
depleted in grains of sizes ranging between $\sim 10^{-2}\,\mu$m and
$2K$.

We tested this scenario by assuming an additional population of
$10^{-2}\,\mu$m compact silicate grains with a surface density
distribution that corresponds to that of the $\beta\dma{pr}=0.45$
particles. The 12\,$\mu$m resolved emission is very well matched in
terms of shape and flux  $\sim$20--30\,AU with a mass of
$10^{-2}\,\mu$m grains of the same order of magnitude as the mass
contained in the particles of size $2K$. The far-IR emission and the
scattered light images, in particular the $r^{-5}$ radial shape of the
surface brightness distribution further than 120\,AU, are not affected
by the presence of such very small grains since they are very poor
scatterers. The presence of such very small grains, and if so, their
chemical composition as well as their size distribution, remain
open issues.
\section{Summary and conclusions}
\label{discuss}
As in \citet{mou97b}, we assumed a disk of planetesimals orbiting
around \bp\ and subject to the gravitational perturbation of a planet
on an inclined orbit.

Assuming that collisions among these planetesimals produce grains with
a size distribution proportional to $a^{-3.5}$ between the blow out
limit and millimeter sizes, and assuming that these grains are subject
to radiation pressure in addition to gravitational forces, and using
plausible chemical composition and optical properties for the grains,
we are able to reproduce the scattered light images (intensity, radial
distribution, warp, ``butterfly'' asymmetry) as well as the far IR
integrated fluxes. The PB are located between 20 and 150\,AU with a
peak at $\sim$100--120\,AU. Due to radiation pressure, the grain
radial distribution strongly depends on their size and the resulting
dust disk extends very far from the star.  The ``butterfly'' asymmetry
is nothing more than the mirror of the warp of PB disk at larger
distances due to radiation pressure.

The 12\,$\mu$m images require an additional population of small grains
close to the star as was evident several years ago and attributed to
comet evaporation. We have shown that radiation pressure acting on
these grains does not produce a distribution that can account for the
observed radial shape of the 12\,$\mu$m emission as found by
\citet{pan97} at distances larger than $\sim$20--30\,AU.  A population
of grains smaller than (or comparable to) the blow-out limit in
addition to the population necessary to reproduce the scattered light
images may account for this $> 30$\,AU 12\,$\mu$m emission. It is
found that the extrapolation of the $a^{-3.5}$ distribution down to
such small sizes is somewhat problematic. A tentative alternative
could be that there is a discrete population of grains with size less
than $10^{-2}\,\mu$m in addition to the main population of grains.
This is plausible as we assume that the main population of grains is
made of porous aggregates ultimately breaking down and producing $<
10^{-2}\,\mu$m grains that could be responsible for the emission
further than 30\,AU at 12\,$\mu$m, without affecting the scattered
light images and the 60--100\,$\mu$m integrated fluxes. As another
alternative, it is also possible to argue that the $a^{-3.5}$ law is
unlikely close to the blow-out limit: grains close to the blow-out
size may be over-abundant simply because of the lack of smaller grains
able to efficiently destroy them. In this latter scenario, a larger
number of small grains is explained as a result of collisions, without
involving additional processes. \\

Finally, our consistent modeling of the \bp ~disk, combining the
various available observations has the following limitations: 1/ the
solution found is probably not unique as various parameters (optical
properties, size distribution, spatial distribution, grain production
mechanisms) are intricately involved; 2/ the application of the
$a^{-3.5}$ size distribution all the way between blow-out size and
parent bodies is questionable; 3/ the side-to-side asymmetry in the
12\,$\mu$m and 20\,$\mu$m images has not yet been explained.
Obviously, this asymmetry originates close to the star and might be
due to the evaporating comets as was proposed by \citet{lec96}.

Further investigations will be able to benefit from new direct
constraints on the spatial distribution of grains, as observed at
various wavelengths (scattered light and thermal emission high angular
resolution images), for the inner disk ($<$ 50 AU), and optical
properties of the grains.
%
%
\begin{acknowledgements}
  We wish to thank Philippe Th\'ebault, Alain Lecavelier, Herv\'e
  Beust for fruitful discussions. J.C. Augereau and A.M. Lagrange
  acknowledge visitor support at QMWC through PPARC grant: PPARC
  GR/J88357.
\end{acknowledgements}

\end{document}